\documentclass[11pt]{article}
\usepackage{amsmath,amsfonts,amssymb}
\usepackage{delarray}

\def\N{\mathbb{N}}

\def\Z{\mathbb{Z}}

\def \id{\mathrm{Id}}
\begin{document}
\begin{center}
{\Large \bf Wick ordering for coherent state quantization in $1+1$ de Sitter space}\\
{\large A. Rabeie$^*$, E. Huguet$^{**}$, J. Renaud$^{***}$}\\
\smallskip

{\small
$^*$~Razi University, \\
Faculty of Science, Bagh-e-Abrisham, Kermanshah, Iran\footnote{rabeie@razi.ac.ir}.\\
$^{**}$~Universit\'e Paris-7-Denis-Diderot, \\APC - Astroparticule et Cosmologie (UMR 7164),
Batiment Condorcet,
10, rue Alice Domon et L\'eonie Duquet  
F-75205 Paris Cedex 13, France\footnote{huguet@apc.univ-paris7.fr}.\\
$^{***}$~Universit\'e Paris-Est, \\APC - Astroparticule et Cosmologie (UMR 7164),
 Batiment Condorcet,
10, rue Alice Domon et L\'eonie Duquet  
F-75205 Paris Cedex 13, France\footnote{jacques.renaud@univ-mlv.fr}.}
\end{center}
\begin{abstract}
 We show that the coherent state  quantization of massive particles in $1+1$ de Sitter space exhibits an 
ordering property: There exist some classical observables $A$ and $A^*$ such that $O_{A^{*p}}O_{A^q}=O_{A^{*p} A^q}$ $p, q \in \Z$, where $O_A$ is the quantum observable corresponding to the classical observable $A$.
\end{abstract}

 \section{Introduction}

A quantization procedure \cite{berezincm} consists in a map $A\to O_A$ which associates an operator (quantum observable) $O_A$ to any function $A$  on phase space (classical observable). Now, when two classical observables are quantized  an additional ordering rule is needed in order to quantize their product. 
As a matter of fact the usual methods of quantization, including the most advanced ones like, for instance, geometric quantization\cite{ali}, allow to quantize only a restricted set of classical observables and do not provide any ordering rule. For instance,  the generators of some Lie 
algebra may be quantized, but even the simplest functions of them  like polynomials  must be considered separately.


On the other hand, coherent states and their generalizations \cite{AliAntoineGazeau} allow to quantize any classical observable. 
In addition, in the Bargman representation, the coherent state (or anti-Wick) quantization corresponds to the following ordering : one puts the $\bar z$ terms on the left and the $z$ terms on the right and then quantizes. At the quantum level this corresponds to the anti-normal ordering in which the derivation terms are on the left. As far as we know, there is no prescription for ordering operators in the general coherent states context, and it could be interesting to exhibit such a rule.

As a matter of fact, and this is the object of the present letter, the coherent states builded
on the 1+1 de Sitter phase space for massive particles \cite{jp} present such an ordering property:
The classical 
observables  which can be expanded as a power series of the two functions
 $A(\beta,J) = e^{\varepsilon J+i\beta}$ and $A^*(\beta,J)$
 are related to the operators $O_{A}$ and $O_{A^*}$ 
which verify $O_{A^{*p}}O_{A^q}=O_{A^{*p} A^q}$ $p, q \in \Z$. It is important to note that these
coherent states are closely related to the coherent states for the motion of massive particle on the
circle \cite{dbg,brs,ok,krp,gdo}.

In sect. 2 we briefly recall the quantization procedure in the framework of 
generalized coherent states. The application to the massive free particle in $1+1$ de Sitter 
space is summarized in sect. 3. The ordering property 
is proved in sect. 4.

\section{General coherent states}
Let $X$ be a set equipped with some
measure $\mu$ and ${\cal
H}$ be a separable sub-Hilbert space of ${\rm L}^2(X,\mu)$. A set of coherent states, and the associated quantization, can be defined if
 there exists a continuous mapping
\begin{equation}\label{xX}
  X \ni x \longrightarrow  | x \rangle \in {\cal H},
\end{equation}
 where
the family of states $\{ | x \rangle \}_{x\in X}$ obeys the following
two conditions:
\begin{enumerate}
\item\label{c1} {\em Normalization} : $\langle \, x\, | x \rangle = 1,$
\item \label{c2}{\em Resolution of the identity in ${\cal H}$}:
 $\int_X  | x\rangle \langle x \, | \, \nu(dx)=
\id_{{\cal H}}$,
\end{enumerate}
 where $\nu(dx)$ is another measure on $X$,
absolutely continuous with respect to
$\mu(dx)$: there exists a positive measurable function $h(x)$ such that $\nu(dx)
= h(x) \mu(dx)$.

In this framework, the coherent states quantization of a {\it classical} observable, that is to say of a
function $f$ on
$X$, consists in associating to  $f$
the operator
\begin{equation}\label{oper}
O_f \equiv  \int_X  f(x) ~ | x\rangle \langle x| \,~
\nu(dx).
\end{equation}
In this context, $f$ is said  to be the upper (or contravariant)
 symbol of the operator $O_f$, whereas the mean value
$\langle x| O_f | x\rangle$ is said to be the lower (or covariant) symbol of
$O_f$ \cite{Perelomov2}. Of course, such a particular quantization scheme is
intrinsically limited to all those classical observables for which
the expansion (\ref{oper}) is mathematically justified within the
theory of operators in Hilbert spaces ({\it e.g.} weak convergence).

In practice, the states $| x\rangle$  can be obtained \cite{jp} from some
superposition of elements of an orthonormal basis $\{ | \phi_n\rangle
\}_{n \in \N}$ of ${\cal H}$ if we assume in addition that
\begin{equation}
{\cal N} (x) \equiv \sum_n \vert \phi_n (x) \vert^2 <
\infty \ \mbox{almost everywhere}.
\label{factor}
\end{equation}
Then, the states
\begin{equation}
| x\rangle \equiv \frac{1}{\sqrt{{\cal N} (x)}} \sum_n
{\phi_n^* (x)}~ | \phi_n\rangle, \label{cs}
\end{equation}
are normalized and satisfy the resolution
of the identity in ${\cal H}$  with 
\begin{equation}
\nu(dx) = {\cal N} (x)\, \mu(dx).
\label{meas}
\end{equation}

\section{Coherent states for a massive particle in $1+1$ dS space}
The above construction is now applied to the phase space of a massive free 
particle in a $1+1$ de Sitter space. This can be realized as follows \cite{jp}. The phase space reads
 \begin{eqnarray}
 X = T^{*}(S^{1})&=&\Big\{(\overrightarrow{x},\overrightarrow{J}) \in \mathbb{R}^{2}\times \mathbb{R}^{2}
 ~~\Big | ~~x^{2}=1, ~\overrightarrow{x}.\overrightarrow{J}=0\Big\}\\
 &=&\Big\{(\beta,J)~~\Big | ~~ 0\leqslant \beta < 2\pi, ~J \in \mathbb{R}\Big\},
 \end{eqnarray}
 where $(\beta,J)$ are conjugate coordinates.
 Let $\mid n \rangle, {n\in \Z}$, be the state vector corresponding to the function
  \begin{equation}
 \Phi_{n}^{\epsilon}(\beta,J)=e^{-\frac{\epsilon n^{2}}{2}}e^{n(\epsilon J+i\beta)},\qquad \qquad n\in
 \mathbb{Z}.
 \end{equation}
 The set $\{|n\rangle, n\in \Z\}$ is an orthonormal family 
of ${\rm L}^2({X,\mu}),$ 
where
 \begin{equation}
 \mu(d\beta,dJ)=\sqrt{\frac{\epsilon}{\pi}}\frac{1}{2\pi}e^{-\epsilon J^{2}}dJd\beta.
 \end{equation}
This family fulfills the condition (\ref{factor}). As a consequence,
 coherent states can be  defined on this phase space through \cite{jp,dbg,brs,ok,krp,gdo}: 
 \begin{equation}
 \mid J,\beta\rangle=\frac{1}{\sqrt{{\cal N}_{\epsilon}(\beta,J)}}\sum_{n}(\Phi_{n}^{\epsilon}
 (\beta,J))^*\mid n \rangle,
 \end{equation}
where ${\cal N}_{\epsilon}(\beta,J)$ is the normalization factor given by the following convergent series ($\epsilon >0$)
 \begin{eqnarray}
 \nonumber {\cal N}_{\epsilon}(\beta,J)={\cal N}_{\epsilon}(J):&=&\sum_{n}\mid \Phi_{n}^{\epsilon}(\beta,J)
 \mid ^{2}\\
 &=&\sum_{n}e^{-\epsilon n^{2}+2\epsilon nJ}<\infty.
 \end{eqnarray}
  These coherent states lead to the following resolution of the identity in ${\cal H}$
 \begin{equation}
 \int_{J=-\infty}^{\infty} \int_{\beta=0}^{2\pi} \mid J,\beta><J,\beta \mid {\cal N}_{\epsilon}(\beta,J)
 ~\mu(d\beta, dJ)=\id_{{\cal H}},
 \end{equation}
 where $\cal H$ is the Hilbert space spanned by the $\mid n \rangle$. We are now in position to quantize classical observables. For instance, applying (\ref{oper})
to $J^m, m \in \N$ and $\beta$ leads to
\begin{eqnarray}
 O_{J^{m}}&=&\sum_{n}
\left( \frac{i}{2\sqrt{\epsilon}}\right)^m
 \mathrm{H}_m(-i\sqrt{\epsilon}n)
 \mid n\rangle\langle n \mid,\label{qua.9}\\
 O_{\beta}&=&\pi \,\id_{{\cal H}} +i \sum_{n\neq n'}\frac{e^{-\frac{\epsilon (n-n')^{2}}{4}}}{n-n'}\mid n\rangle\langle n'
 \mid ,\label{qua.6}
 \end{eqnarray}
where $\mathrm{H}_m$ is the Hermite polynomial.

 \section{Ordering algebra}\label{ordering}
 
 It is well known \cite{Perelomov2} that the contravariant quantization, in the context of the standard coherent states, corresponds to the anti-normal ordering. More precisely, in the Bargman representation $O_z=z$ and $O_{z^*}=\frac{\partial}{\partial z}$. Thus the upper symbol associated with the classical observable  $f = \sum A_{nm}(z^*)^nz^m$
reads:
 \begin{equation}\label{Of}
 O_f = \sum A_{nm}\left(\frac{\partial}{\partial z}\right)^n z^m.
 \end{equation}
 Something very similar appears in our context.
Let $f$ be a classical observable which admits the series representation $\sum_{p,q}c_{p,q}A^{*p} A^q$ 
$p, q \in \Z$, where
$ A=e^{+\varepsilon J+i\beta}$.
Then, $f$ can be quantized as~:
\begin{equation}\label{mainres}
O_f=\sum_{p,q}c_{pq}O_{A^{*p}}O_{ A^q},~~~p, q \in \Z,
\end{equation}
where the $A^*$'s appear on the leftmost position. 
This is the main result of this letter. The proof is easy, here it is: 
matrix elements of operators associated to $A^{*p}$,  $A^q$ and $A^{*p}A^q$ 
are calculated using (\ref{oper}). It follows that 
\begin{eqnarray*}
O_{A^{*p}}&=&\sum_ne^{\frac{\varepsilon}{2} p(p + 2n)}|n\rangle\langle n+p|,\\
O_{A^q}&=&\sum_ne^{-\frac{\varepsilon}{2} q(q - 2n)}|n\rangle\langle n-q|,\\
O_{A^{*p} A^q}&=&\sum_{n}e^{\frac{\varepsilon}{2}(2pq - (p+q)(q - p - 2n))}|n\rangle\langle n-(q-p)|.
\end{eqnarray*}
A straightforward calculation shows that
\begin{equation}\label{result} 
O_{A^{*p}}O_{A^q}=O_{A^{*p} A^q},
\end{equation}
from which (\ref{mainres}) follows using linearity. Note that 
\begin{eqnarray*}
O_{A^*}&=& \left(O_A \right)^\dagger,\\
O_{A^{-1}}&=&\left(O_A \right)^{-1}.
\end{eqnarray*}

The operator (\ref{mainres}) is weakly defined as soon as $p$ and $q$ admit a lower bound. Indeed, the matrix 
elements of (\ref{mainres}) read
\begin{equation*}
\langle m \mid O_f \mid n \rangle = \sum_{\substack{p,q\\p+q = m - n}} c_{pq} e^{\frac{\varepsilon}{2} 
(p(p - 2m) + q(2n + q))},
\end{equation*} 
which is a finite sum in this case.


\end{document}